\documentclass[amsmath,amssymb,twocolumn,aps,prd,10pt]{revtex4-2}
\usepackage[T1]{fontenc}
\usepackage{graphicx}
\usepackage{subfig}
\usepackage{epsfig}
\usepackage{dcolumn}
\usepackage{multirow}
\usepackage{rotating}
\usepackage{verbatim}
\usepackage{bm}
\usepackage{color}
\usepackage{soul}
\usepackage{float}
\usepackage{hyperref}
\usepackage{booktabs}

\def\lsim{\raise0.3ex\hbox{$<$\kern-0.75em\raise-1.1ex\hbox{$\sim$}}}
\def\gsim{\raise0.3ex\hbox{$>$\kern-0.75em\raise-1.1ex\hbox{$\sim$}}}

\def\beqa{\begin{eqnarray}}
\def\eeqa{\end{eqnarray}}

\begin{document}

\title{Exclusive $Z^0$ production in $ep$ and $eA$ collisions at high energies}
\author{G. M. Peccini} 
\email{guilherme.peccini@ufrgs.br}
\author{M. V. T. Machado}
\email{magnus@if.ufrgs.br}
\affiliation{High Energy Physics Phenomenology Group, GFPAE. Universidade Federal do Rio Grande do Sul (UFRGS)\\
Caixa Postal 15051, CEP 91501-970, Porto Alegre, RS, Brazil} 
\author{L. S. Moriggi}
\email{lucasmoriggi@unicentro.br}
\affiliation{Universidade Estadual do Centro-Oeste (UNICENTRO), Guarapuava, PR, Brazil}

\begin{abstract}
In this work the $k_{\perp}$-factorization formalism is applied to compute the exclusive $Z^0$ boson  photoproduction in $ep$ and $eA$ collisions. The study is also extended to $pp$ and $AA$ processes. The nuclear effects are investigated considering heavy and light ions. Analytical models for the unintegrated gluon distribution are taken into account and the corresponding theoretical uncertainty is quantified. The analysis is done for electron-ion collisions at the Large  Hadron-Electron Collider (LHeC), its high-energy upgrade (HE-LHeC) and at the Future Circular Collider (FCC) in lepton-hadron mode. Additionally, ultra-peripheral heavy ion collisions at future runs of the Large Hadron Collider (LHC) and at the FCC (hadron-hadron mode) are also considered.

\end{abstract}

\maketitle

\section{Introduction} 

The Large Hadron Collider (LHC) has studied physics at TeV scale allowing the access to unexplored  kinematical regimes at large luminosities. This machine is able to search and study the
physics beyond of the Standard Model (SM) with high accuracy. Traditionally, a baseline SM  signal is the $Z^0$ production \cite{2017JPhG...44b3001B}. In proton-proton ($pp$) collisions, the hadronic $Z^0$ decays are not easy to identify due to the strong background of QCD multijet production from hadronic event environment \cite{phdthesis}. The  high-statistics measurements in final states with leptons are the main channel at the LHC. On the other hand, the exclusive production of $Z^0$ at electron-ion colliders or in ultra-peripheral collisions (UPCs) present some advantages. There is an increasing interest on exclusive processes at the LHC. \cite{Albrow:2010yb,Harland-Lang:2015cta}. One of these favorable conditions is  the processes to be perturbatively calculable with not so large uncertainties due to the high mass of the boson. Another feature is the clear experimental signature compared to the $Z^0$ signal coming from hadroproduction.  A dedicated experimental search  has already been done for $p\bar{p}$ collisions at the Tevatron \cite{CDF:2009xwv}. No exclusive $Z\rightarrow \ell^+\ell^-$  candidates were observed leading to the first upper limit on the exclusive $Z^0$ cross section, $\sigma (p\bar{p}\rightarrow pZ^0\bar{p}) < 0.96$ pb. Similar searches have been preliminarily carried out at the LHC \cite{MedinaJaime:2015nri}. Therefore, it is timely to investigate the prediction for both electron-ion colliders and ultra-peripheral collisions given their high energies and integrated luminosities (see Ref. \cite{Hentschinski:2022xnd} where prospects for exclusive processes which are complementary between the LHC and the Electron-Ion Collider - EIC are discussed).

The exclusive $Z^0$ photoproduction  in electron-proton collisions was first addressed in the pioneering work of Ref. \cite{Bartels:1981jh}, where an analysis using non-forward QCD planar ladder diagrams was done and applied to  diffractive (inclusive and semi-inclusive) boson production in $ep$ colliders. The simple two-gluon exchange model of the Pomeron was used in Ref. \cite{Pumplin:1996pi} to compute the boson photoproduction cross section. There,  a finite gluon mass has been included in the propagators to suppress the long distance contributions. In the context of the color dipole picture, the exclusive $Z^0$ production has been analysed in both space-like \cite{Goncalves:2007vi} and time-like kinematics \cite{Motyka:2008ac}. The equivalent calculation in $k_{\perp}$-factorization approach was presented first in Ref. \cite{Cisek:2009hp}. In Refs. \cite{Motyka:2008ac,Cisek:2009hp}, applications to photoproduction in $pp$ collisions at the LHC energies were performed. The analysis for UPCs in $pp$, $pA$ and $AA$ collisions at the LHC has been done in Ref. \cite{COELHO2020115013}.

In this paper, the  main goal is to compute the exclusive $Z_0$ production cross section in $ep$ and $eA$ collisions by using the $k_{\perp}$-factorization formalism. Different models for the unintegrated gluon distribution (UGDs), ${\cal{F}}(x,k_{\perp}^2)$,  will be considered. We focus on the energies and phase space of the Large Electron-Hadron Collider (LHeC) \cite{LHeCStudyGroup:2012zhm,LHeC:2020van,Andre:2022xeh} and the Future Circular Collider (FCC) \cite{FCC:2018byv,FCC:2018vvp} in $eh$ mode (FCC-eh). Using the obtained cross section for $Z^0$ production in photon-proton and photon-nucleus processes, the corresponding predictions for $pp$ and $AA$ collisions are computed. In the last case, the cross section at the energies of the High Luminosity LHC (HL-LHC), High Energy LHC (HE-LHC) and FCC are calculated. The sources of theoretical uncertainties are investigated. An important point to be highlighted is that this present paper extends our previous works on the exclusive dilepton production in lepton-hadron and hadron-hadron machines \cite{Peccini:2020jkj,Peccini:2021rbt}. This work is organized as follows. In Section \ref{theorysection}, the theoretical formalism for $ep$ (subsection \ref{secep}) and $eA$ collisions (subsection \ref{seceA} ) is briefly reviewed along with the calculations concerning the analytical models for UGDs in proton and nuclei. In Section \ref{numresults}, the numerical results are presented for the energies relevant to the LHeC machine. The number of events per year is obtained without imposing any further kinematic cuts. In addition, the cross section for photoproduction in $pp$ collisions and in  $AA$ UPCs are studied using the  equivalent photon approximation. A comprehensive analysis is performed for light nuclei and lead as well. In the last section, we summarize the results and present the main conclusions.

\begin{table}[t]
\begin{center}
\begin{tabular}{|l|c|c|c|}
\hline 
Collider & $E_e$ (GeV) & $E_p$ (TeV) & $\sqrt{s}$ (TeV) \\
\hline
LHeC/HL-LHeC    & 60 & 7 &  1.3 \\
HE-LHeC & 60 & 13.5 &  1.7 \\
FCC-eh     & 60 & 50 &  3.5 \\
\hline
\end{tabular}
\end{center}
\caption{Energies of the beams at future electron-proton colliders (LHeC/HL-LHeC, HE-LHeC and FCC-eh).}
\label{tab:1}
\end{table}

\section{Theoretical formalism}
\label{theorysection}

\subsection{Exclusive Z$^0$ production in electron-proton collisions}
\label{secep}

The calculation of exclusive Z$^0$ production cross section follows the same formalism of TCS (timelike Compton scattering) \cite{Peccini:2020jkj,Peccini:2021rbt,Schafer:2010ud}. In the context of the $k_T$-factorization approach it was first proposed in Ref. \cite{Cisek:2009hp} using only one model for the UGD. Afterwards, in Ref. \cite{Peccini:2020jkj} the present authors computed the TCS cross section for dilepton production in $ep$ collisions by using different and updated  UGDs. Specifically, four UGDs containing distinct physical information were analyzed. Moreover, in Ref. \cite{Peccini:2021rbt} we have accounted for dilepton production via TCS in electron-nucleus collisions assuming the UGD proposed in Ref. \cite{Moriggi:2020qla} (named as MPM hereafter). In the present work, we will consider the same UGDs utilized in Ref. \cite{Peccini:2020jkj}  to evaluate the cross section for exclusive Z$^0$ production.   

Comparing to TCS, one can calculate the cross section for exclusive $Z^0$ boson production by simply replacing the electromagnetic photon-quark coupling by the electroweak one, $e e_f \to \frac{e g_V^f}{sin 2 \theta_W}$, where $\theta_W$ is the Weinberg angle. Only the weak vector coupling is relevant, where $g_V^{f}=(I_3^f -2e_f \sin^2 \theta_W)/ \sin 2 \theta_W$) \cite{Motyka:2008ac}. The weak isospin of a quark of flavour $f$ and charge $ee_f$ is $I_3^f$. Along with the coupling replacement, one has also to redefine $x$ in terms of the $Z^0$ mass, 
\begin{eqnarray}
x=\xi_{\mathrm{sk}} \left(\frac{M_Z^2}{W^2}\right),
\end{eqnarray}
where $\xi_{\mathrm{sk}}$ is inserted in order to correct the skewedness effect \cite{Schafer:2010ud}. Following Ref. \cite{Cisek:2009hp}, the value $\xi_{\mathrm{sk}} = 0.41$ has bee considered. 
\begin{table}[t]
\begin{center}
\begin{tabular}{|l|c|c|c|}
\hline 
Nucleus & LHeC/HL-LHeC  & HE-LHeC  & FCC-eA   \\
\hline
O  & 0.92 & 1.27 &  2.45 \\
Ar & 0.87 & 1.21 &  2.32 \\
Kr & 0.85 & 1.18 &  2.27 \\
Pb & 0.81 & 1.13 &  2.18 \\
\hline
\end{tabular}
\end{center}
\caption{Center-of-mass energies (in units of TeV) at future electron-nucleus colliders (LHeC/HL-LHeC, HE-LHeC and FCC-eA) for different nuclei.}
\label{tab:2}
\end{table}

Taking the equation for the imaginary part of TCS amplitude expressed in Refs. \cite{Peccini:2020jkj,Peccini:2021rbt,Schafer:2010ud} and performing the coupling replacement, one can obtain the forward amplitude for exclusive $Z^0$ production in $ep$ collisions \cite{Cisek:2009hp}:
\begin{eqnarray}
\label{eq:ImAf}
{\cal{M}}(W,|t|=0)& = & \sum_{f}\frac{2W^2\alpha_{em}g_V^f}{\pi} \nonumber \\
&\times & \int_0^1 dz\int d^2\vec{\kappa}_{\perp}\frac{(i+\rho_R)\mathrm{Im}A_f(z,\vec{\kappa}_{\perp})}{\kappa_{\perp}^2+ m_f^2 - z(1 - z)M_Z^2 - i\varepsilon}, \nonumber \\
\end{eqnarray}
where $\rho_R$ is the ratio of real to imaginary part of amplitude. Moreover, $m_f$ is the quark mass of flavor $f$ and $W^2$ is the center-of-mass energy squared of the photon-proton system. The quantity $\mathrm{Im}A_f$ is defined as
\begin{eqnarray}
\mathrm{Im}A_f (z,\vec{\kappa}_{\perp}) = \int \frac{\pi dk_{\perp}^2}{k_{\perp}^4}\alpha_s (\mu^2){\cal{F}}(x,k_{\perp}^2)\,C(z,\kappa_{\perp},k_{\perp},m_f), \nonumber
\end{eqnarray}
where the explicit expression for the function $C(z,\kappa_{\perp},k_{\perp},m_f)$ can be found in Refs. \cite{Cisek:2009hp,Peccini:2020jkj}. The following hard scale $\mu^2= \mathrm{max}(\kappa_{\perp}^2+m_f^2,k_{\perp}^2)$ has been chosen.  The corresponding amplitude for the $\gamma p \rightarrow Z^0p$ process within the diffraction cone is written as
\begin{eqnarray}
{\cal{M}}(W,|t|) = {\cal{M}}(W,|t|=0)e^{-B_D|t|},
\end{eqnarray}
where the energy dependent diffraction slope, $B_D$, is parametrized as $B_D=B_0+2\alpha^{\prime}_{\mathrm{eff}} \log (W^2/W_0^2)$. Here,  $\alpha^{\prime}_{\mathrm{eff}}=0.164$ GeV$^{-2}$, $B_0=3.5$ GeV$^{-2}$ and $W_0=95$ GeV  \cite{Cisek:2009hp}. 

\begin{table*}[t]
\caption{Cross section in units of fb and event rates/year times branching ratio  for exclusive $Z^0$ photoproduction in $ep$ and $eA$ collisions. The results are presented for the MPM UGD model as a baseline. Numerical calculation are presented for O and Pb nuclei.}
\label{tab:3}
\begin{tabular}{|c|c|c|c|}
\hline
Collider & Nucleus & $\sigma_{ep(A)}$ (fb) & Number of events per year \\ \hline
HL-LHC   & p       & $7.11$              & 60.6                      \\ 
         & O       & $70.3$              & 3.28                     \\ 
         & Pb      & $1.97 \times 10^3$  & 7.07       \\ \hline
HE-LHC   & p       & $10.9$              & 140                       \\ 
         & O       & $113$               & 13.69                       \\ 
         & Pb      & $3.27 \times 10^3$  & 30.09        \\ \hline
FCC      & p       & $30.9$              & 494     \\ 
         & O       & $349$               & 125.62    \\ 
         & Pb      & $1.04 \times 10^4$  & 287.98   \\ \hline
\end{tabular}
\end{table*}

The differential and the integrated production cross section are, respectively, given by
\begin{eqnarray}
{\frac{d \sigma}{dt}}(\gamma p \to Z^0 p) & = &\frac{ \left|{\cal{M}}(W,|t|)\right|^2}{16  \pi }  ,\\
\sigma (\gamma p \to Z^0 p)  & = & \frac{ \big [\mathrm{Im} ({\cal{M}}^{\gamma p \to Z^0 p }) \big ] ^2 \left(1+\rho_R^2  \right)}{16  \pi B_D}  ,
\label{eq:xsecZ0}
\end{eqnarray}

The cross section in Eq. (\ref{eq:xsecZ0}) will be evaluated for the LHeC (as well for its high luminosity and high energy upgrades) \cite{LHeCStudyGroup:2012zhm,LHeC:2020van,Andre:2022xeh,Agostini:2020fmq} and for the FCC center-of-mass energies \cite{FCC:2018byv,FCC:2018vvp}. These energies are summarized in Table \ref{tab:1}. Concerning the quark flavours, $u,\,d\,,s\,,c,\,b$ are considered. 

For the numerical calculation, three models for the UGD will be taken into consideration: the Moriggi-Peccini-Machado (MPM) \cite{Moriggi:2020zbv}, Ivanov-Nikolaev (IN) \cite{Ivanov:2000cm} and Golec-Biernat-Wusthoff (GBW) \cite{GolecBiernat:1998js,Golec-Biernat:2017lfv} models. The MPM and GBW models are analytical, whereas IN is dependent on the input for the DGLAP evolved gluon distribution. The first one present also the geometric scaling property, i.e., the UGD depends on the scaling function $\tau_s=k_{\perp}^2/Q_{sp}^2(x)$. The proton saturation scale in this case scales on $x$ in the form $Q_{sp}^2=(x_0/x)^{\lambda}$. The three models are parametrized as follows:
\begin{eqnarray}
{\cal{F}}_{\mathrm{MPM}} (x,k_{\perp}^2) &= &  k_{\perp}^2 \frac{3 \sigma_0}{4 \pi ^2 \alpha_s}    \frac{(1+\delta n)\tau_s} {(1+ \tau_s)^{2+\delta n}}, \\
{\cal{F}}_{\mathrm{GBW}} (x,k_{\perp}^2) &= & k_{\perp}^2 \frac{3 \sigma_0}{4 \pi ^2 \alpha_s}  \tau_s e^{-\tau_s},\\
{\cal{F}}_{\mathrm{IN}} (x,k_{\perp}^2) &= &{\cal{F}}_{\mathrm{soft}}^{(B)}(x,k_{\perp}^2) \frac{\kappa_s^2}{k_{\perp}^2+ \kappa_s^2} \nonumber \\
&+& {\cal{F}}_{\mathrm{hard}}(x,k_{\perp}^2) \frac{k_{\perp}^2}{k_{\perp}^2+ \kappa_h^2}.
\end{eqnarray}

In the MPM, one has $\delta n = a \tau ^b$ and $Q_s^2=(x_0/x)^{0.33}$ (here, $\lambda =0.33$ value is fixed), where  the parameters $\sigma_0$, $x_0$, $a$ and $b$ were fitted against DIS data in the kinematic domain $x<0.01$ \cite{Moriggi:2020zbv}. A fixed value $\alpha_s=0.2$ has been considered. Beside describing DIS data at small-$x$, it also describes the spectra of produced hadrons in $pp$ and $p\bar{p}$ processes. This model was built by means of the geometric scaling approach and a Tsallis-like behavior of the measured spectra. For the GBW parametrization, the updated parameters  $\sigma_0$, $x_0$ and $\lambda$ (fit including bottom quark contribution) are taken from Ref. \cite{Golec-Biernat:2017lfv}. Concerning the IN model, the functions ${\cal{F}}_{\mathrm{sof}}^{(B)}$ and ${\cal{F}}_{\mathrm{hard}}$ as well as the quantities $\kappa_{s,h}$ can be found in Ref. \cite{Ivanov:2000cm}.

In what follows, we will perform the extension of the above approach to $eA$ collisions. The dependence on energy and atomic number is addressed as well as the estimation of the theoretical uncertainty.

\subsection{Exclusive Z$^0$ production in electron-nucleus collisions}
\label{seceA}

In case of electron-nucleus collisions, the main difference from electron-proton collisions is the unintegrated gluon distribution. Accordingly, we will utilize a nuclear UGD instead of a proton UGD here. The nuclear UGD considered will be obtained by applying the Glauber-Mueller formalism. Additionally, one has to take into account the nuclear form factor, $F_A(q)$. In this work, an analytic form factor given by a hard sphere of radius $R_A =1.2A^{1/3}$ fm, convoluted with a Yukawa potential with range $a = 0.7$ fm, has been considered  \cite{PhysRevC.14.1977}. Therefore, the differential cross section for $eA$ collisions is given by
\begin{eqnarray}
{\frac{d \sigma}{dt}}(\gamma A \to Z^0 A)  & = &  \frac{ \big [\text{Im} ({\cal{M}}^{\gamma A \to Z^0 p }) \big ] ^2 (\left(1+\rho_R^2  \right) }{16  \pi } |F_A(q)|^2, \,\,\,\,\,\,\, \,\\
F_A(q) & = &  \frac{4\pi\rho_0}{A |q^3|} \left( \frac{1}{1+a^2q^2} \right)  \nonumber \\
&\times & \left[ \sin{(q R_A)} - q R_A\cos{(q R_A)}  \right], 
\end{eqnarray}
where $q=\sqrt{|t|}$. Namely, the amplitude depends on $t$ in a factorized way,  ${\cal{M}}(W^2,t)={\cal{M}}(W^2,t=0) F_A(q)$.

For $Z^0$ production in nuclear collisions, we will investigate the nuclei proposed in the LHC prospects (see Ref. \cite{Citron:2018lsq,Bordry:2018gri,Amoroso:2022eow}), namely O, Kr, Ar  and Pb. The energy of the nuclear beams are given by the energy of the proton beam multiplied by the ratio $Z/A$, where $Z$ is the atomic number while $A$ is the atomic mass number. In Table \ref{tab:2} , we outline the beam energies along with the center-of-mass energies for electron-nucleus collisions. The energy of the electron beam is $60$ GeV.

As stressed out previously, the nuclear unintegrated gluon distribution is in order rather than the proton one to compute the amplitude in Eqs. \ref{eq:ImAf}. In this context, in Ref. \cite{Moriggi:2020qla} the MPM model was adapted to nuclear targets by using Glauber-Gribov formalism \cite{Gribov:1968jf,Gribov:1968gs}, leading to the following form for the nuclear UGD:
\begin{equation}
\label{eq:fiA}
{\cal{F}}_A(x,k_{\perp}^2,b)=k_{\perp}^2\frac{3}{4\pi^2\alpha_s}k_{\perp}^2\nabla^2_{k_{\perp}} \mathcal{H}_0 \left\{ \frac{1-S_{dip}^A(x,r,b)}{r^2}\right\},
\end{equation}
where $\mathcal{H}_0 \left\{ f(r) \right\}=\int dr\, r J_0(k_{\perp}r)f(r)$ is the order zero Hankel transform and the quantity $S_{dip}^A$ is given by \cite{Armesto:2002ny}
\begin{equation}
\label{eq:SdA}
S_{dip}^A(x,r,b)=\mathrm{e}^{-\frac{1}{2} T_A(b)\sigma_{dip}(x,r)} \ .  
\end{equation}
Here, $\sigma_{dip}(x,r)$ is the dipole cross section for the proton case. The function $T_A(b)$ is the thickness function and depends on the impact parameter, $b$, and normalization $\int d^2b T_A(b) = A $. A Woods--Saxon parametrization for the nuclear density \cite{DEVRIES1987495} has been considered.

\begin{figure}[t]
\centering
    \includegraphics[width=0.5\textwidth]{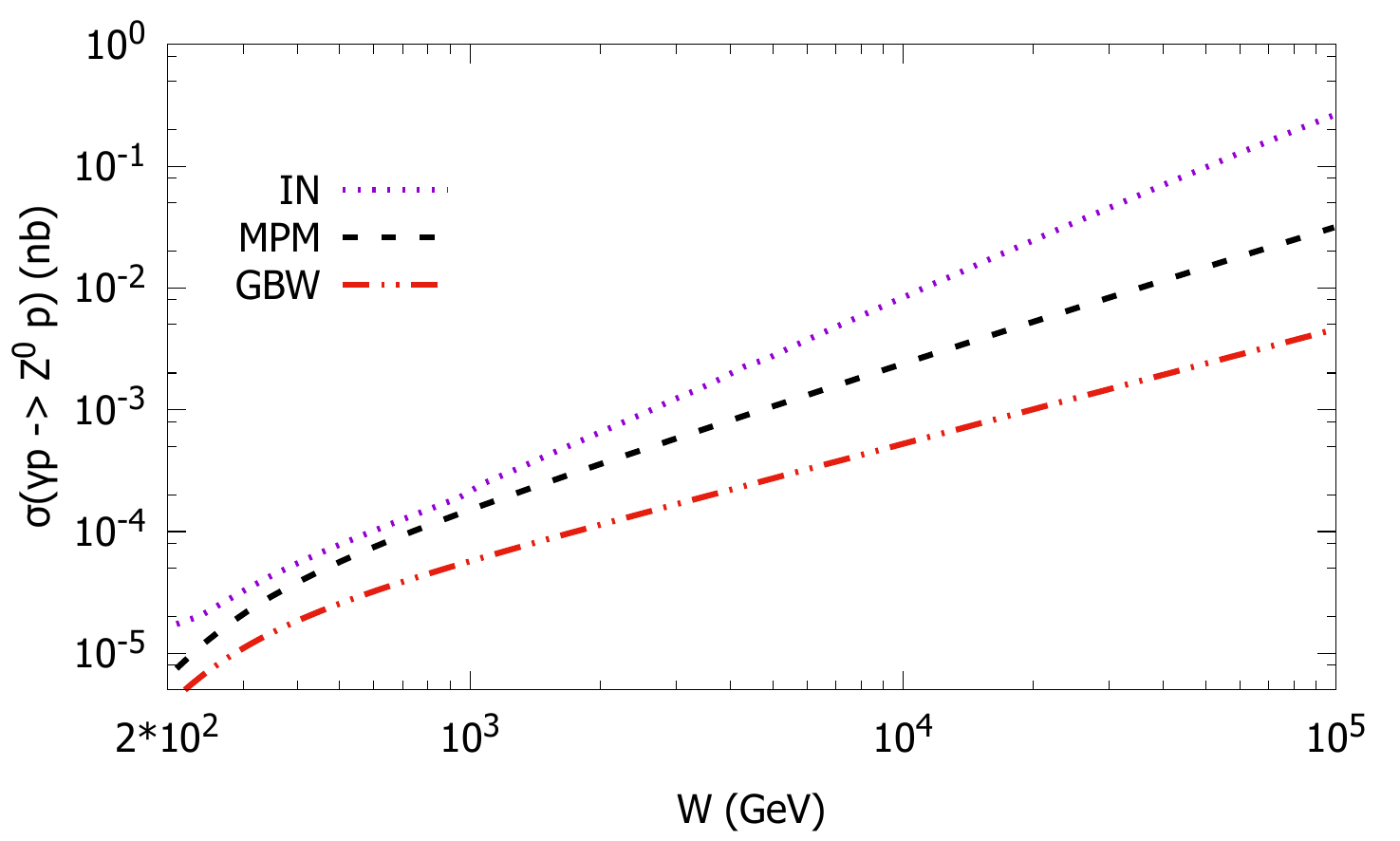}
    \caption{Cross section $\sigma(\gamma^* p \rightarrow Z^0 p)$ as a function of photon-proton center of mass energy, $W_{\gamma p}$. Numerical results for IN, MPM and GBW phenomenological models are presented.}
    \label{fig:epZ}
\end{figure}

In case of a proton target, a homogeneous object with radius $R_p$ is assumed which factorizes  $S_{dip}(x,r,b)$ into $S_{dip}(x,r,b)=S_{dp}(x,r)\Theta (R_p-b)$. For large dipoles, $S_{dip}(x,r) \rightarrow 0$ and the cross section reaches a bound given by $\sigma_0=2\pi R_p^2$. Within the saturation approach, the gluon distribution presents a maximum at $k_{\perp} \simeq Q_s(x)$. The dipole cross section in coordinate space $r$ for MPM model may be evaluated as \cite{Moriggi:2020zbv},
\begin{eqnarray}
\label{eq:DISc}
\sigma_{dip}(\tau_r)=\sigma_0\left (  1-\frac{2(\frac{\tau_r}{2})^{\xi}K_{\xi}(\tau_r)}{\Gamma(\xi)} \right ),
\end{eqnarray}
where $\xi =1+\delta n$ and $\tau_r = rQ_s(x)$ is the scaling variable in the position space. Accordingly, the nuclear gluon distribution is obtained from Eqs. (\ref{eq:fiA}) and (\ref{eq:SdA}). The same procedure has been applied to the IN and GBW models. Interestingly, as the hard scale associated to the process is $\mu^2=m_Z^2$, one expects that small dipoles (large $k_{\perp}$ gluons) will be the dominant contribution to the cross section. This means that $\mu^2 \gg Q_{s,A}(x)^2$ and the nuclear shadowing should be quite small. A good approximation for the nuclear UGD would be ${\cal{F}}_A(x,k_{\perp}) \approx A{\cal{F}}_p(x,k_{\perp})$.  

Models for nuclear UGDs are very scarce in literature. It would be worth comparing the present calculations with the numerical results from nuclear UGDs evolved by DGLAP or CCFM evolution equations as studied in Refs. \cite{deOliveira:2013oma,Modarres:2019ndk,Modarres:2018ymh}. The advantage would be the introduction of other effects as anti shadowing and EMC effects. The models based on Glauber-Gribov formalism bring only information on the shadowing effects to the referred process. Another source of theoretical uncertainty is the treatment for the skwedness correction once the effect is enhanced in the production amplitude squared.

\begin{figure*}[t]
\centering
    \includegraphics[width=0.8\textwidth]{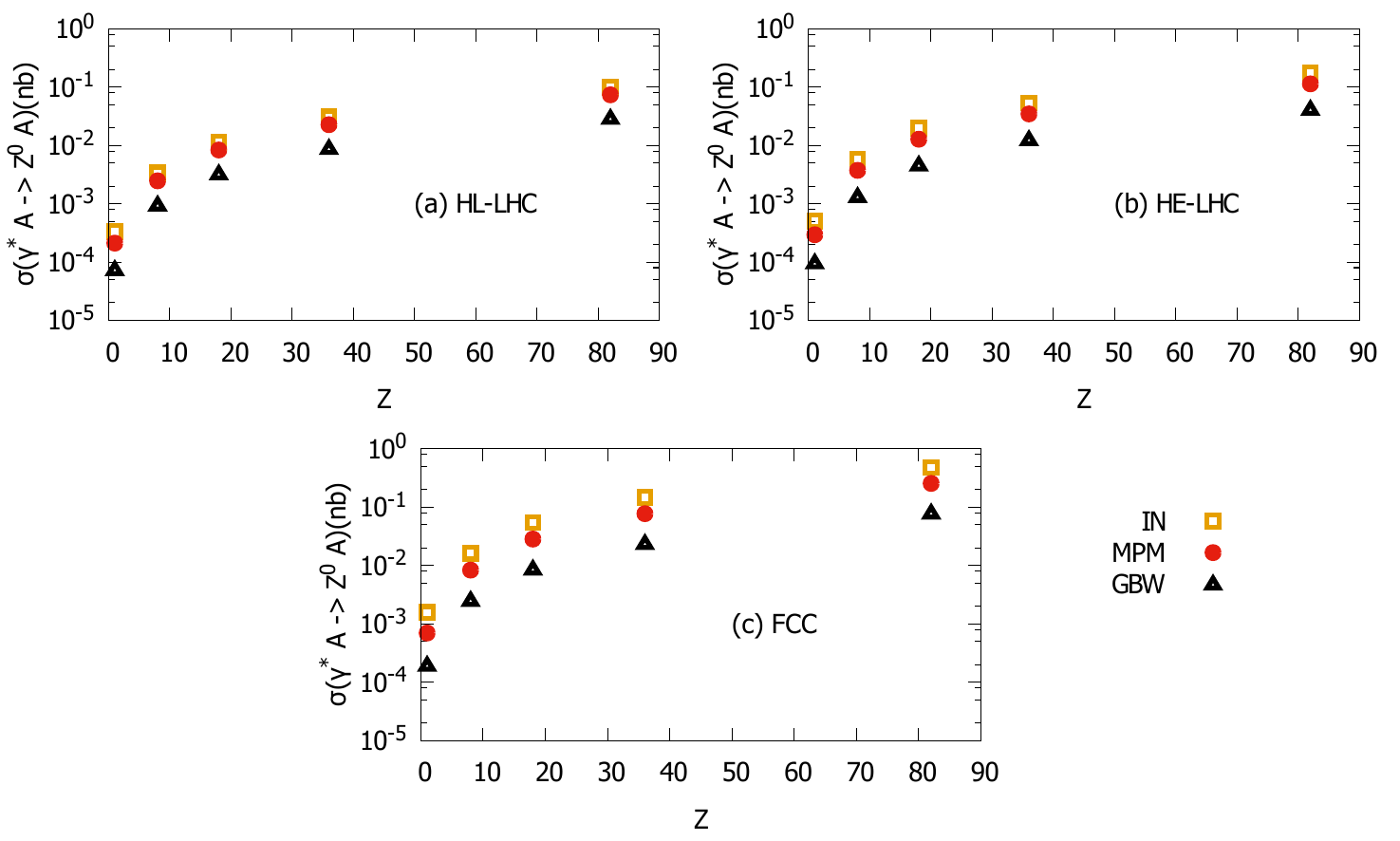}
    \caption{Cross section  $\sigma(\gamma^* p(A) \rightarrow Z^0 p(A))$ as a function of atomic number $Z$ at the energies of (a) HL-LHC, (b) HE-LHC and (c) FCC. The predictions from IN, MPM and GBW models are shown for O, Ar, Kr, Pb nuclei and proton as well. The corresponding energies are presented in Tab. \ref{tab:2}.}
    \label{fig:eAZ}
\end{figure*}

\section{Numerical results and discussion}
\label{numresults}

\begin{table*}[t]
	\caption{The event rates/year for exclusive $Z^0$ photoproduction in $pp$ and PbPb collisions in different rapidity ranges. The results are presented for the MPM UGD model as a baseline. }
	\label{tab:4}
	\setlength{\tabcolsep}{8pt}
	\centering
	\begin{tabular}{ccccc}
		\toprule %
		\textbf{Collider} & \multicolumn{2}{c}{\textbf{pp collisions}} & \multicolumn{2}{c}{\textbf{PbPb collisions}} \\ \cmidrule(lr){1-1}\cmidrule(lr){2-3}\cmidrule(lr){4-5}
		\centering
		\small
		\textbf{} & \small$-2.0<y<+2.0$ & \small$+2.0<y<+4.5$ & \small $-2.0<y<+2.0$  & \small $+2.0<y<+4.5$ \\
		\midrule %
		HL-LHC  &  $1.39\times 10^3$ & 531 & 364 & 45.8 \\
		HE-LHC  & $8.15\times 10^3$ & 461 & $1.33\times 10^3$ & 183  \\
		FCC   & $7.54\times 10^3$ & $6.60\times 10^4$ & $5.92\times 10^4$ & $8.31\times 10^3$ \\
		\bottomrule %
	\end{tabular}
\end{table*}

Let us start by presenting the results for the $Z^0$ photoproduction in $\gamma p$ scattering. The corresponding energies for the $ep$ colliders are exhibited in Table \ref{tab:1}.  In Fig.  \ref{fig:epZ} the predictions for the IN (dotted curve), MPM (dashed curve) and GBW (dot-dashed curve) models are shown as a function of photon-proton center of mass energy, $W_{\gamma p}$. In the TeV energy scale the cross section has the order of magnitude of $\sigma (\gamma^* p \rightarrow Z^0p)\approx 0.1$ pb and a large  theoretical uncertainty. Accordingly, the GBW model gives a lower bound for the cross section values and weaker energy behavior compared to MPM and IN. The reason is the DGLAP-like evolution embedded in both IN and MPM models for the UGD. The  $x$ value probed at $W_{\gamma p }=1$ TeV is $\sim 10^{-3}$. The output coming from the MPM model can be parametrized in the following way:  $\sigma_{\mathrm{MPM}} (\gamma p \rightarrow Z^0p)= [180\, \mathrm{fb}] \,(W_{\gamma p}/W_0)^{1.13}$ (with $W_0=10^3\,\mathrm{GeV}$). Notice that the prediction from Ref. \cite{Cisek:2009hp} is properly reproduced here by using the Ivanov-Nikolaev UGD. A steeper growth is predicted by Motyka and Watt (MW) in Ref. \cite{Motyka:2008ac}, where the color dipole picture is considered and by using the impact parameter saturation model (IP-SAT) and timelike $Z^0$ boson. The IP-SAT parametrization includes DGLAP evolution for the dipole cross section and the result scales as  $\sigma_{\mathrm{MW}} (\gamma p \rightarrow Z^0p)= [37\, \mathrm{fb}] \,(W_{\gamma p}/W_0)^{1.73}$, with $W_0=1.3\times 10^3\,\mathrm{GeV}$.

\begin{figure*}[t]
\centering
    \includegraphics[width=0.8\textwidth]{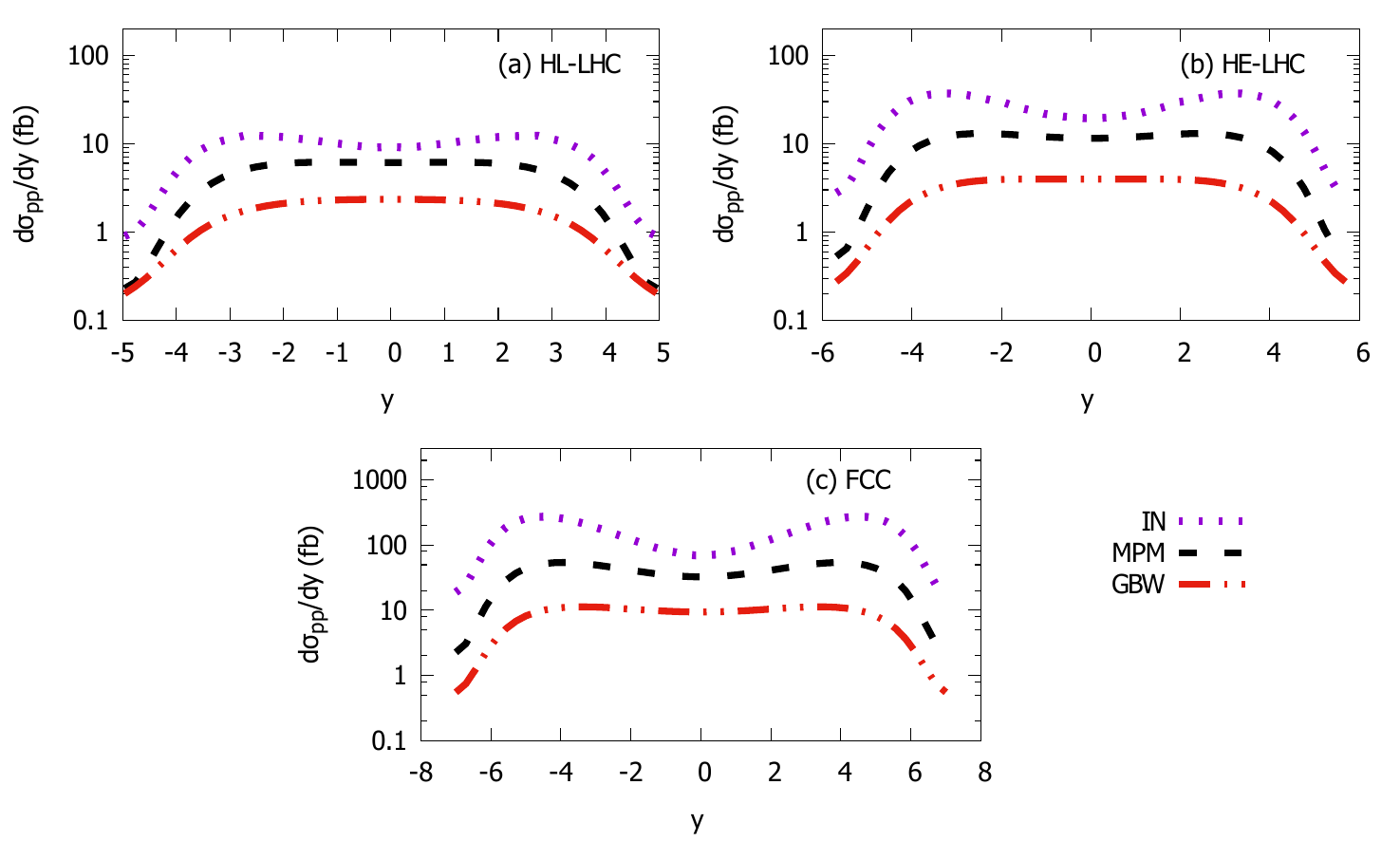}
    \caption{Rapidity distribution for exclusive $Z^0$ production in $pp$ collisions. Predictions are shown for the energies of (a) HL-LHC, (b) HE-LHC and (c) FCC. }
    \label{fig:ppZ}
\end{figure*}
The analyses for nuclear targets are presented in Fig.  \ref{fig:eAZ}. Predictions from the three phenomenological models are shown for the nuclear species  presented in Table \ref{tab:2} and for proton as a baseline. As examples of order of magnitude one has $\sigma (\gamma Pb \rightarrow Z^0Pb)\approx 84.6\,(260)$ pb  and $\sigma (\gamma O \rightarrow Z^0O)\approx 2.46\,(8.3)$ pb at HL-LHC(FCC) energy. The dependence on atomic mass number from MPM model (for $A>1$) is given by $\sigma_{\mathrm{MPM}}^{\gamma A}=\sigma_A\,A^{\delta}$, where $\sigma_A = 50.5$ fb and $\delta = 1.39$ at the HL-LHC and $\sigma_A = 216$ fb and $\delta = 1.33 $ at the FCC. This result is consistent with the weak absorption limit for the nuclear dipole cross section typical for $Z^0$ production.  In the figure the predictions are shown for $W_{\gamma A}^{max} = \sqrt{s_{eA}}$. 

In Table \ref{tab:3} the cross section times branching ratio into dileptons (in units of fb) is presented for $ep$ and $eA$ collisions. Here, the interest is in very small $Q^2\ll 1$  GeV$^2$ range where the photoproduction cross section is independent of photon virtuality. Therefore,  the $ep(A) \rightarrow eZ^0p(A)$ cross section can be written as
\begin{eqnarray}
\frac{d\sigma}{dW^2} & = &\frac{\alpha_{em}}{2\pi s}\left[ \frac{1+(1-y)^2}{y}\ln \frac{Q_{max}^2}{Q_{min}^2 } - \frac{2(1-y)}{y}\left(1- \frac{Q_{min}^2}{Q_{max}^2 }  \right) \right]\nonumber \\
&\times &\sigma^{\gamma p(A)}(W^2),
\end{eqnarray}
where $y$ is the inelasticity variable and $Q_{min}^2 = m_e^2y^2/(1-y)$. 
The corresponding number of events per year is also presented. The run with Oxigen is comparable in number of events for the proton target. The experimental feasibility is enhanced in $eA$ case compared to the $ep$ machine. This is specially important when kinematic cuts are imposed in order to remove the main dilepton QED background. The predictions considering the decay into hadrons is larger by a factor 20 but the  experimental feasibility worsens.

\begin{figure*}[t]
\centering
    \includegraphics[width=0.8\textwidth]{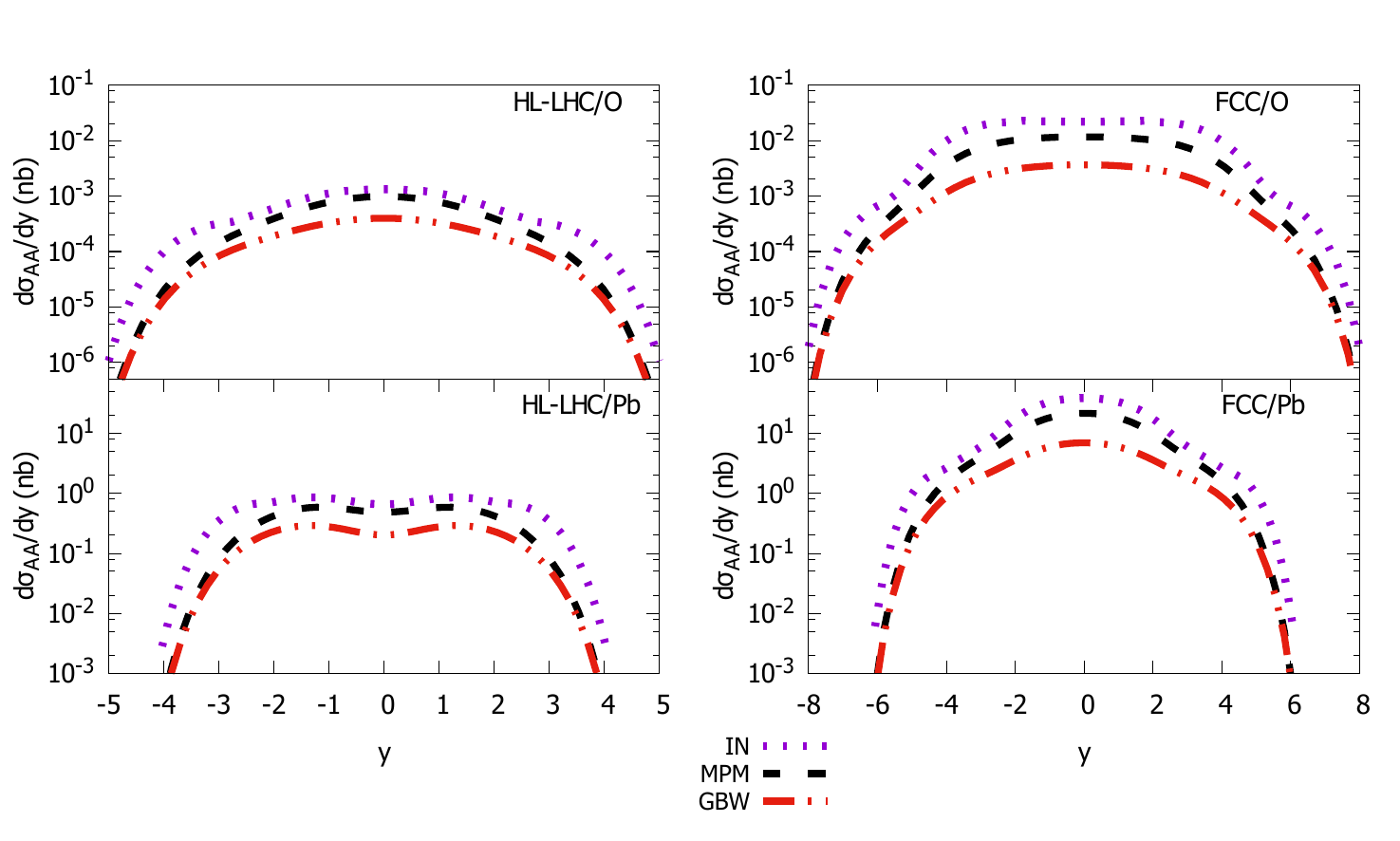}
    \caption{Rapidity distribution for $Z^0$ production in  $AA$ collisions. Prediction are presented for Oxigen (upper plots) and Lead (lower plots) nuclei at the energies of HL-LHC (plots on the left) and FCC (plots on the right).}
    \label{fig:AAZ}
    \end{figure*}
    
Let us now move to ultra-peripheral collisions. The cross section to produce a $Z^0$ boson  in a proton-proton collision within the Weisz\"acker-Williams approximation is given by \cite{Klein:1999qj,Klein:2020fmr,Klein:2016yzr}
\begin{eqnarray}
\sigma (p p \rightarrow  p p + Z^0 ) = 2\int_{0}^{\infty} \frac{dn_{\gamma}^p}{d\omega}\sigma (\gamma +p\rightarrow Z^0+p) \,d\omega , \nonumber
\end{eqnarray}
where $\omega $ is the photon energy and $dn_{\gamma}^p/d\omega$ is the photon spectrum for protons. In the numerical calculations we have used the photon spectrum from Ref. \cite{Drees:1988pp}. The corresponding rapidity distribution is obtained as follows:
\begin{eqnarray}
\frac{\sigma (pp \rightarrow  pp+Z^0 p)}{dy} = \omega \frac{dn_{\gamma}^p}{d\omega}\sigma_{\gamma +p\rightarrow Z^0+p} (\omega), 
\end{eqnarray}
in which the rapidity $y$ of the produced $Z^0$ state with mass $M_Z$ is
related to the photon energy through $y = \ln(2\omega/M_Z)$. The rapidity distributions
are shown in Fig. \ref{fig:ppZ} and the the calculations are for collision
energies of (a) the HL-LHC, (b) HE-LHC and (c) FCC colliders. The dotted, dashed and dot-dashed curves are the results for the IN, MPM and GBW UGDs, respectively. Here, the predictions are presented  without absorption effects which depend on the rapidity. For instance, the absorptive correction at 14 TeV for $y=0$ is $\langle S^2\rangle \simeq 0.8$ whereas it is $\langle S^2\rangle \simeq 0.6$ for $y=2$ \cite{Cisek:2009hp}. The theoretical uncertainty is still sizable. At the energy of HL-LHC, our predictions are in agreement with those in Refs. \cite{Motyka:2008ac,Cisek:2009hp,COELHO2020115013}. In general, the numerical results obtained using $k_T$-factorization approach are higher than those from color dipole framework. The rapidity distribution for higher hadron energies (HE-LHC and FCC) can be directly compared with the results of Ref. \cite{COELHO2020115013}. There, two dipole cross section have been considered (bCGC and IP-SAT models) within the color dipole picture. The order of magnitude of the cross sections are in agreement. The experimental feasibility is promising by using the dilepton decay channel. A careful analysis by using kinematic cuts should remove the large background coming from $\gamma \gamma \rightarrow \ell^+\ell^-$ process. The search for exclusive $Z^0$ production in proton-proton collisions can follow similar methodology employed in the corresponding search in $p\bar{p}$ collisions at Tevatron energies \cite{CDF:2009xwv}.
\begin{figure*}[t]
\centering
    \includegraphics[width=0.8\textwidth]{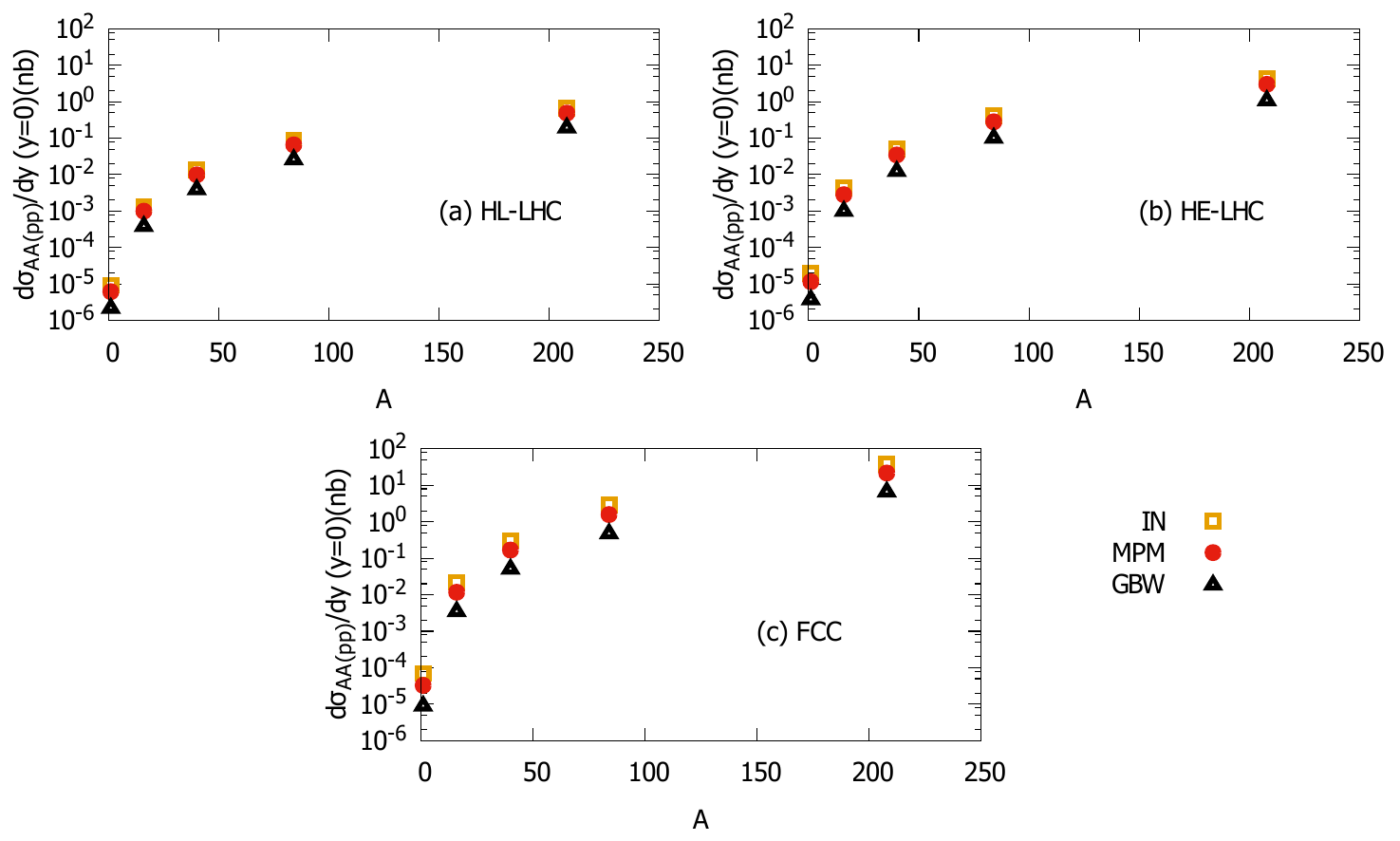}
    \caption{Rapidity distribution for $Z^0$ production in  $AA$ collisions at mid-rapidity $y=0$ as a function of the atomic mass number $A$. The predictions from IN, MPM and GBW models are shown for O, Ar, Kr, Pb nuclei and proton as well.  Predictions are shown for the energies of (a) HL-LHC, (b) HE-LHC and (c) FCC. }
    \label{fig:AAZy0}
\end{figure*}

Finally, the $Z^0$ exclusive production is investigated in ultra-peripheral heavy ion collisions (UPCs). The corresponding rapidity distribution for the coherent production is given by \cite{Bertulani:2005ru}
\begin{eqnarray}
\frac{\sigma (AA \rightarrow  AZ^0 A)}{dy} = \omega \frac{dn_{\gamma}^A}{d\omega}\sigma_{\gamma +A\rightarrow Z^0+A} (\omega), 
\end{eqnarray}
where $dn_{\gamma}^A/d\omega$ is the photon spectrum for nuclei. The analytical photon flux  for $b>2R_A$ has been used in \cite{Klein:1999qj}. In Fig.  \ref{fig:AAZ} the rapidity distribution is shown for Oxigen (O) and Lead (Pb) nuclei: upper and lower plots, respectively. This is presented for the energies of HL-LHC and FCC. The predictions are somewhat larger than the ones presented in Ref. \cite{COELHO2020115013} where the color dipole approach is considered. There, $d\sigma_{AA} (y=0)/dy\simeq 0.6$ nb in contrast with present calculation $d\sigma_{AA} (y=0)/dy\simeq 10$ nb. In both approaches the theoretical uncertainty is large, specially that one associated with the specific model for the dipole cross section or UGD.

In Fig. \ref{fig:AAZy0} the predictions for rapidity distributions at mid-rapidity, $y=0$, are shown for the nuclear species presented in Table \ref{tab:2} as well as for protons. The cross sections are exhibited as a function of the atomic mass number $A$ for the energies of (a) HL-LHC, (b) HE-LHC and (c) FCC. The predictions for MPM and IN models are quite similar at this scale and a lower bound is given by the GBW model.

In order to summarize the results for $pp$ and PbPb collisions in Table \ref{tab:4} the event rates/year are presented, where the production cross section has been multiplied by branching ratio for decays into  dileptons.  Two rapidity ranges are considered: $|y|\leq 2$ (central rapidities) and $+2.0\leq y\leq +4.5$. (forward rapidities). Results are presented for the MPM UGD model as a representative example of application. The present calculation is more comprehensive than those in Ref. \cite{COELHO2020115013} since light nuclei are also taken into account.

\section{Conclusions}

In this work the exclusive production of $Z^0$ boson is investigated in $ep$ and $eA$ collisions within the  $k_T$-factorization formalism. The theoretical uncertainty is studied by comparing the results for different unintegrated gluon distributions available in literature. It was found that the corresponding variance is large when models containing parton saturation effects are contrasted to those where they are not applied. The analysis is done in the kinematic range of interest of EIC and the LHeC. As a by product the $Z^0$ photoproduction is also investigated in $pp$ and $AA$ collisions. The application was restricted to the coherent scattering and predictions for incoherent scattering would be valuable.  A comprehensive study is done concerning different nuclear species relevant for the LHC future runs.  The experimental measurement feasibility is briefly discussed.

\label{summary}

\section*{Acknowledgments}

This work was financed by the Brazilian funding
agency CNPq. We are thankful to Krzysztof Piotrzkowski (Université Catholique de Louvain) for his contribution in this paper.

\bibliographystyle{h-physrev}
\bibliography{referencias_z0}

\begin{thebibliography}{10}

\bibitem{2017JPhG...44b3001B}
J.~{Berryhill} and A.~{Oh},
\newblock Journal of Physics G Nuclear Physics {\bf 44}, 023001 (2017).

\bibitem{phdthesis}
R.~Chislett,
\newblock {\em Studies of hadronic decays of high transverse momentum W and Z
  bosons with the ATLAS detector at the LHC},
\newblock PhD thesis, University College London, 2014.

\bibitem{Albrow:2010yb}
M.~G. Albrow, T.~D. Coughlin, and J.~R. Forshaw,
\newblock Prog. Part. Nucl. Phys. {\bf 65}, 149 (2010), 1006.1289.

\bibitem{Harland-Lang:2015cta}
L.~A. Harland-Lang, V.~A. Khoze, and M.~G. Ryskin,
\newblock Eur. Phys. J. C {\bf 76}, 9 (2016), 1508.02718.

\bibitem{CDF:2009xwv}
CDF, T.~Aaltonen {\em et~al.},
\newblock Phys. Rev. Lett. {\bf 102}, 222002 (2009), 0902.2816.

\bibitem{MedinaJaime:2015nri}
M.~Medina~Jaime,
\newblock {\em {Produ\c{c}\~ao exclusiva de b\'osons Z em colis\~oes pp no
  experimento CMS/LHC}},
\newblock PhD thesis, Campinas State U., 2015.

\bibitem{Hentschinski:2022xnd}
M.~Hentschinski {\em et~al.},
\newblock (2022), 2203.08129.

\bibitem{Bartels:1981jh}
J.~Bartels and M.~Loewe,
\newblock Z. Phys. C {\bf 12}, 263 (1982).

\bibitem{Pumplin:1996pi}
J.~Pumplin,
\newblock (1996), hep-ph/9612356.

\bibitem{Goncalves:2007vi}
V.~P. Goncalves and M.~V.~T. Machado,
\newblock Eur. Phys. J. C {\bf 56}, 33 (2008), 0710.4287,
\newblock [Erratum: Eur.Phys.J.C 61, 351 (2009)].

\bibitem{Motyka:2008ac}
L.~Motyka and G.~Watt,
\newblock Phys. Rev. D {\bf 78}, 014023 (2008), 0805.2113.

\bibitem{Cisek:2009hp}
A.~Cisek, W.~Schafer, and A.~Szczurek,
\newblock Phys. Rev. D {\bf 80}, 074013 (2009), 0906.1739.

\bibitem{COELHO2020115013}
R.~Coelho and V.~Gonçalves,
\newblock Nuclear Physics B {\bf 956}, 115013 (2020).

\bibitem{LHeCStudyGroup:2012zhm}
LHeC Study Group, J.~L. Abelleira~Fernandez {\em et~al.},
\newblock J. Phys. G {\bf 39}, 075001 (2012), 1206.2913.

\bibitem{LHeC:2020van}
LHeC, FCC-he Study Group, P.~Agostini {\em et~al.},
\newblock J. Phys. G {\bf 48}, 110501 (2021), 2007.14491.

\bibitem{Andre:2022xeh}
K.~D.~J. Andr\'e {\em et~al.},
\newblock Eur. Phys. J. C {\bf 82}, 40 (2022), 2201.02436.

\bibitem{FCC:2018byv}
FCC, A.~Abada {\em et~al.},
\newblock Eur. Phys. J. C {\bf 79}, 474 (2019).

\bibitem{FCC:2018vvp}
FCC, A.~Abada {\em et~al.},
\newblock Eur. Phys. J. ST {\bf 228}, 755 (2019).

\bibitem{Peccini:2020jkj}
G.~M. Peccini, L.~S. Moriggi, and M.~V.~T. Machado,
\newblock Phys. Rev. D {\bf 102}, 094015 (2020), 2010.03101.

\bibitem{Peccini:2021rbt}
G.~M. Peccini, L.~S. Moriggi, and M.~V.~T. Machado,
\newblock Phys. Rev. D {\bf 103}, 054009 (2021), 2101.08338.

\bibitem{Schafer:2010ud}
W.~Schafer, G.~Slipek, and A.~Szczurek,
\newblock Phys. Lett. B {\bf 688}, 185 (2010), 1003.0610.

\bibitem{Moriggi:2020qla}
L.~S. Moriggi, G.~M. Peccini, and M.~V.~T. Machado,
\newblock (2020), 2012.05388.

\bibitem{Agostini:2020fmq}
LHeC, FCC-he Study Group, P.~Agostini {\em et~al.},
\newblock (2020), 2007.14491.

\bibitem{Moriggi:2020zbv}
L.~Moriggi, G.~Peccini, and M.~Machado,
\newblock (2020), 2005.07760.

\bibitem{Ivanov:2000cm}
I.~Ivanov and N.~N. Nikolaev,
\newblock Phys. Rev. D {\bf 65}, 054004 (2002), hep-ph/0004206.

\bibitem{GolecBiernat:1998js}
K.~J. Golec-Biernat and M.~Wusthoff,
\newblock Phys. Rev. D {\bf 59}, 014017 (1998), hep-ph/9807513.

\bibitem{Golec-Biernat:2017lfv}
K.~Golec-Biernat and S.~Sapeta,
\newblock JHEP {\bf 03}, 102 (2018), 1711.11360.

\bibitem{PhysRevC.14.1977}
K.~T.~R. Davies and J.~R. Nix,
\newblock Phys. Rev. C {\bf 14}, 1977 (1976).

\bibitem{Citron:2018lsq}
Z.~Citron {\em et~al.},
\newblock CERN Yellow Rep. Monogr. {\bf 7}, 1159 (2019), 1812.06772.

\bibitem{Bordry:2018gri}
F.~Bordry {\em et~al.},
\newblock (2018), 1810.13022.

\bibitem{Amoroso:2022eow}
S.~Amoroso {\em et~al.},
\newblock (2022), 2203.13923.

\bibitem{Gribov:1968jf}
V.~N. Gribov,
\newblock Sov. Phys. JETP {\bf 29}, 483 (1969).

\bibitem{Gribov:1968gs}
V.~N. Gribov,
\newblock Zh. Eksp. Teor. Fiz. {\bf 57}, 1306 (1969).

\bibitem{Armesto:2002ny}
N.~Armesto,
\newblock Eur. Phys. J. C {\bf 26}, 35 (2002), hep-ph/0206017.

\bibitem{DEVRIES1987495}
H.~{De Vries}, C.~{De Jager}, and C.~{De Vries},
\newblock Atomic Data and Nuclear Data Tables {\bf 36}, 495  (1987).

\bibitem{deOliveira:2013oma}
E.~de~Oliveira, A.~Martin, F.~Navarra, and M.~Ryskin,
\newblock JHEP {\bf 09}, 158 (2013), 1307.2825.

\bibitem{Modarres:2019ndk}
M.~Modarres and A.~Hadian,
\newblock Nucl. Phys. A {\bf 983}, 118 (2019), 1901.07772.

\bibitem{Modarres:2018ymh}
M.~Modarres and A.~Hadian,
\newblock Phys. Rev. D {\bf 98}, 076001 (2018), 1901.06477.

\bibitem{Klein:1999qj}
S.~Klein and J.~Nystrand,
\newblock Phys. Rev. C {\bf 60}, 014903 (1999), hep-ph/9902259.

\bibitem{Klein:2020fmr}
S.~Klein and P.~Steinberg,
\newblock (2020), 2005.01872.

\bibitem{Klein:2016yzr}
S.~R. Klein, J.~Nystrand, J.~Seger, Y.~Gorbunov, and J.~Butterworth,
\newblock Comput. Phys. Commun. {\bf 212}, 258 (2017), 1607.03838.

\bibitem{Drees:1988pp}
M.~Drees and D.~Zeppenfeld,
\newblock Phys. Rev. D {\bf 39}, 2536 (1989).

\bibitem{Bertulani:2005ru}
C.~A. Bertulani, S.~R. Klein, and J.~Nystrand,
\newblock Ann. Rev. Nucl. Part. Sci. {\bf 55}, 271 (2005), nucl-ex/0502005.

\end{thebibliography}

\end{document}